\documentclass{article}

\usepackage[sglblindworkshop, final]{ai4nextg_neurips_2025}

\usepackage[utf8]{inputenc} 
\usepackage[T1]{fontenc}    
\usepackage{hyperref}       
\usepackage{url}            
\usepackage{booktabs}       
\usepackage{amsfonts}       
\usepackage{nicefrac}       
\usepackage{microtype}      
\usepackage{xcolor}         
\usepackage{nicefrac}       
\usepackage{amsmath}
\usepackage{mathtools}
\usepackage{amssymb}
\usepackage{multirow}
\usepackage{enumerate}
\usepackage{epsfig}
\usepackage{afterpage}
\usepackage[caption=false,font=footnotesize]{subfig}
\usepackage{lipsum}
\usepackage{textcomp}
\usepackage{algorithm, algorithmic}
\usepackage{setspace}
\usepackage{tabularx}
\usepackage{subfig}
\usepackage{verbatim}
\usepackage{textcomp}
\usepackage{etoolbox}
\usepackage{multicol}
\usepackage{subcaption} 
\usepackage{graphicx}

\title{Adaptive Cooperative Transmission Design for Ultra-Reliable Low-Latency Communications via Deep Reinforcement Learning}

\author{%
  Hyemin~Yu \\
  University of Victoria\\
  Victoria, BC V8P 5C2, Canada \\
  \texttt{hmyu@uvic.ca} \\
   \And
   Hong-Chuan~Yang \\
   University of Victoria \\
  Victoria, BC V8P 5C2, Canada \\
  \texttt{hy@uvic.ca} \\
}

\begin{document}

\maketitle

\begin{abstract}
Next-generation wireless communication systems must support ultra-reliable low-latency communication (URLLC) service for mission-critical applications. 
Meeting stringent URLLC requirements is challenging, especially for two-hop cooperative communication. 
In this paper, we develop an adaptive transmission design for a two-hop relaying communication system. 
Each hop transmission adaptively configures its transmission parameters separately, including numerology, mini-slot size, and modulation and coding scheme, for reliable packet transmission within a strict latency constraint. 
We formulate the hop-specific transceiver configuration as a Markov decision process (MDP) and propose a dual-agent reinforcement learning-based cooperative latency-aware transmission (DRL-CoLA) algorithm to learn latency-aware transmission policies in a distributed manner. 
Simulation results verify that the proposed algorithm achieves the near-optimal reliability while satisfying strict latency requirements. 
\end{abstract}

\section{Introduction} 
Next-generation wireless communication systems are expected to support a wide range of mission-critical applications, such as remote surgery, autonomous vehicles, and real-time virtual/augmented reality \cite{URLLC_intro_0}. 
These use cases demand ultra-reliable and low-latency communication (URLLC) with packet error rates as low as $10^{-5}$ or even $10^{-7}$ and an end-to-end latency on the order of milliseconds \cite{URLLC_intro_1}. 
However, it is challenging to satisfy such stringent URLLC requirements over wireless channels due to unpredictable channel fading and limited radio resources.

Cooperative communication has emerged as a promising solution to enhance transmission reliability in wireless networks by introducing an intermediate relay node between source and destination \cite{URLLC_relaying}. 
However, most existing works on two-hop transmission under latency requirements focus on one-shot transmission with no retransmissions \cite{FBL_relaying_1}, \cite{FBL_relaying_2}, \cite{FBL_relaying_3}. 
Accordingly, any decoding error on either hop leads to transmission failure. 
Moreover, these one-shot schemes have assumed perfect knowledge of the channel state information (CSI) on both hops to optimally allocate channel uses across two communication links. 
Acquiring such global CSI requires excessive overhead and thus cannot comply with the tight latency budget of URLLC. 
The automatic repeat request (ARQ) protocols can enhance reliability via per-hop retransmissions without requiring global CSI \cite{Relaying_ARQ_1}. 
Although effective, ARQ-based retransmission mechanisms inevitably increase transmission delay \cite{URLLC_relaying_latency_1}, \cite{URLLC_relaying_latency_2}. 
Therefore, achieving URLLC in relay-aided transmission requires a novel design that jointly optimizes both reliability and latency rather than sacrificing one for the other.

The Third Generation Partnership Project (3GPP) has introduced 5G new radio (NR) with key features, including adaptive modulation and coding (AMC), scalable numerology, and mini-slot scheduling to support URLLC services \cite{3GPP}. 
Previous works have optimized these features separately, focusing solely on AMC \cite{AMC_learning_based} or on scalable numerology \cite{5G_NR_numerology_2}, which limits the full potential of 5G NR. 
Very recently, Saatchi \textit{et al.} demonstrated significant improvements in reliability by jointly optimizing numerology, mini-slot size, and modulation and coding scheme (MCS) under stringent latency constraints in a point-to-point single‐carrier transmission system \cite{URLLC_adpative_trans_1}, which was extended to multicarrier transmission \cite{URLLC_adpative_trans_2}. 
To the best of our knowledge, however, there has been no existing work that considers the impact of ARQ-based retransmission on the reliability under the latency constraint for two-hop relaying transmission. 

Motivated by this, in this paper, we fill this gap by optimally configuring transmission parameters in every (re)transmission attempt on two hops to maximize the probability of successful packet delivery within a latency budget. 
Given the stringent latency constraint of URLLC, the resource configuration at the transceiver is performed only based on local CSI.  
To enable distributed operation without global CSI exchange/estimation, we propose a dual-agent reinforcement learning-based cooperative latency-aware transmission (DRL-CoLA) algorithm, where the source and the relay act as agents to learn latency-aware transmission policies from local observations and ARQ feedback. 
By doing so, the DRL-CoLA algorithm enables decentralized hop-specific execution while aligning both agents with the end-to-end latency constraint, thereby achieving URLLC over two-hop relaying transmission. 

\section{Two-Hop Transmission Under Latency Constraint} 
\begin{figure}[t]
\centering
\includegraphics[width=0.7\columnwidth]{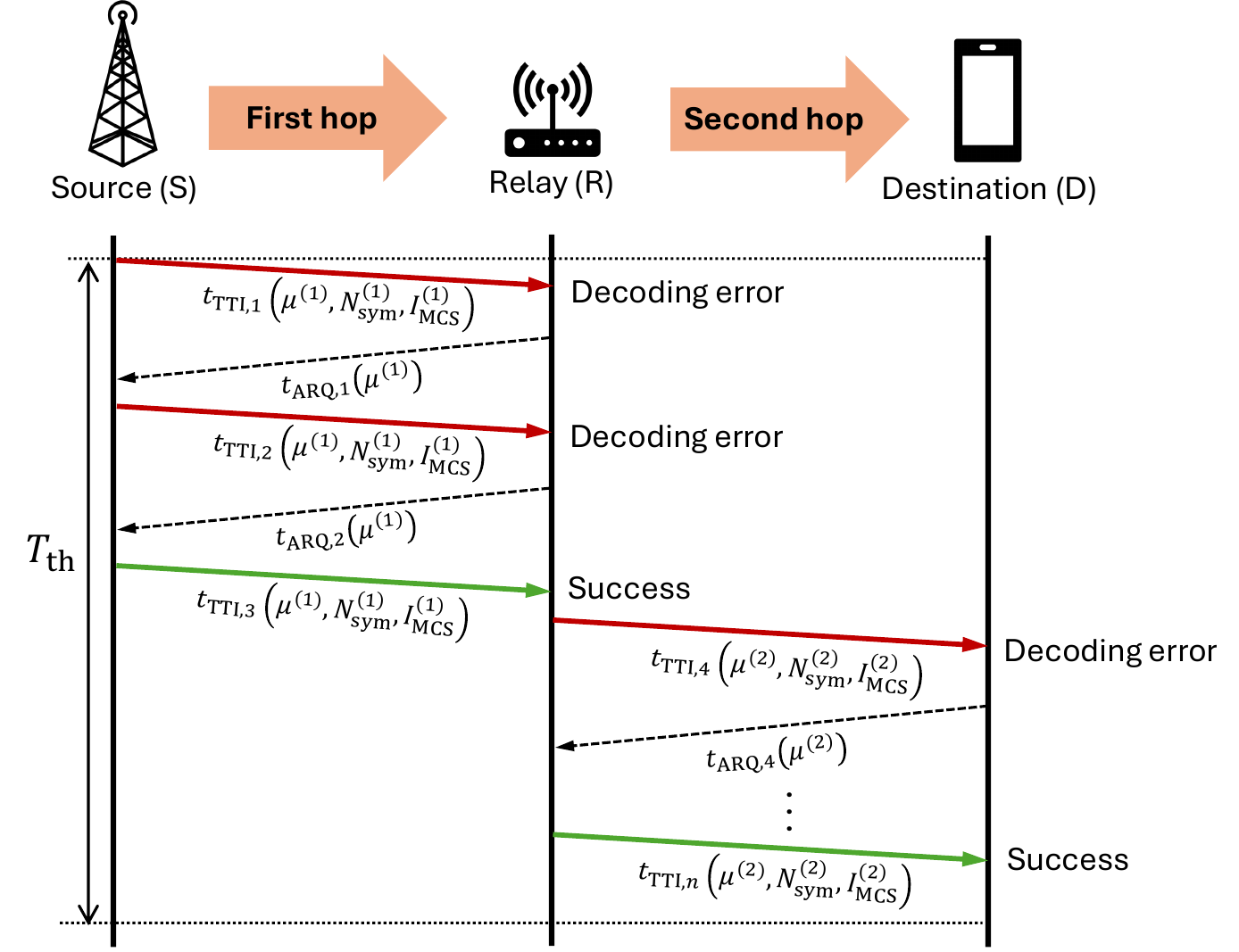}
\caption{System model for two-hop relaying with ARQ protocols under the latency constraint $T_{\rm th}$. }
\label{fig:system_model}
\end{figure}

As depicted in Fig. \ref{fig:system_model}, we consider a two-hop cooperative communication over 5G NR, where a source (S) transmits a delay-sensitive packet to a destination (D) via a half-duplex relay (R). 
To improve reliability, R helps forward a packet from S to D in a decode-and-forward manner, where only a successfully decoded packet at R is forwarded to D. 
We assume that the direct transmission from S to D is unavailable due to severe path loss, deep fading, and/or obstacles. 
Thus, the transmission from S to D via R will perform over two sequential hops, each of which supports scalable numerology, variable mini-slot, and AMC.

As specified in \cite{3GPP_numerology}, 5G NR supports scalable numerology $\mu$ to adjust the subcarrier spacing by $ 2^{\mu} \times 15$ kHz, where $\mu \in \{0, 1, 2, 3, 4\}$. 
The mini-slot transmission in 5G NR further reduces latency by allowing $N_{\rm sym} \in \{2, 4, 7, 14 \}$ orthogonal frequency division multiplexing (OFDM) symbols per mini-slot. 
The selection of numerology $\mu$ and the mini-slot size $N_{\rm sym}$ affects the subframe length, which is given in milliseconds (ms) by $T_{\rm sf} = N_{\rm sym} / (14 \times 2^{\mu})$ \cite{URLLC_adpative_trans_1}. 
To improve link reliability, S and R learn to adapt the MCS to the measured signal-to-noise ratio (SNR) and the remaining latency budget. 
We index the MCS for URLLC use cases specified in \cite{3GPP} by $I_{\rm MCS}$, which is detailed in Table \ref{tab:1}. 
The data packet of $H$ bits must be successfully delivered to D within a latency budget $T_{\rm th}$. 
To this end, S and R select numerology $\mu$, mini-slot size $N_{\rm sym}$, and MCS $I_{\rm MCS}$ for each (re)transmission attempt. 
Each hop transmission is performed over $N_{\rm sc}$ subcarriers, which is given by $N_{\rm sc} = \left \lfloor W /(2^{\mu} \times 15 \times 10^{3})  \right \rfloor$, where $W$ is the available bandwidth, and $ \lfloor . \rfloor$ is the floor function. 

\begin{table*}[t]
\centering
\caption{MCS index table for URLLC service \cite{3GPP}}
\setlength{\tabcolsep}{6pt} 
\renewcommand{\arraystretch}{1.3} 
\scriptsize 
\begin{tabular}{|c||cccccccc|ccc|cccc|}
\hline
$I_{\rm MCS}$ & \multicolumn{1}{c|}{\ 1 \ }  & \multicolumn{1}{c|}{ \ 2 \ }  & \multicolumn{1}{c|}{ \ 3 \ }  & \multicolumn{1}{c|}{4}   & \multicolumn{1}{c|}{5}   & \multicolumn{1}{c|}{6}   & \multicolumn{1}{c|}{7}   & 8   & \multicolumn{1}{c|}{9}   & \multicolumn{1}{c|}{10}  & 11  & \multicolumn{1}{c|}{12}  & \multicolumn{1}{c|}{13}  & \multicolumn{1}{c|}{14}  & 15  \\ \hline \hline
$R_{C} \times 1024$ & \multicolumn{1}{c|}{  30 } & \multicolumn{1}{c|}{ 50} & \multicolumn{1}{c|}{ 78} & \multicolumn{1}{c|}{120} & \multicolumn{1}{c|}{193} & \multicolumn{1}{c|}{308} & \multicolumn{1}{c|}{449} & 602 & \multicolumn{1}{c|}{378} & \multicolumn{1}{c|}{490} & 616 & \multicolumn{1}{c|}{466} & \multicolumn{1}{c|}{567} & \multicolumn{1}{c|}{666} & 772 \\ \hline
Modulation          & \multicolumn{8}{c|}{QPSK $(M=2)$}                                                                                                                                                               & \multicolumn{3}{c|}{16QAM $(M=4)$}                          & \multicolumn{4}{c|}{64QAM $(M=6)$}                                                     \\ \hline
\end{tabular}
\label{tab:1}
\end{table*}

As the latency requirements for URLLC applications are typically shorter than the channel coherence time \cite{FBL_relaying_1}, we model the wireless channels between any two nodes as quasi-static flat fading, where all subcarriers experience the same fading within the latency constraint $T_{\rm th}$. 
Let $h_{1}$ and $h_{2}$ denote the channel coefficients of S-R and R-D links, respectively, which are assumed to follow Rayleigh fading with unit mean $\mathbb{E}\big[ |h_{1}|^{2} \big] = \mathbb{E}\big[ |h_{2}|^{2} \big]=1$. 
Accordingly, the instantaneous SNRs on the S–R and R–D links $\gamma_{1}$ and $\gamma_{2}$ are independent and exponentially distributed with means $\bar{\gamma}_{1} \triangleq P_{1} d^{-\eta}_{1}/ \sigma^{2} $ and $\bar{\gamma}_{2} \triangleq P_{2} d^{-\eta}_{2}/ \sigma^{2} $, where $P_{1}$ and $P_{2}$ are the transmit power at S and R, respectively, $d_{1}$ and $d_{2}$ are the respective link distances, $\eta$ is the path loss exponent, and $\sigma^{2}$ is the power of the additive white Gaussian noise (AWGN).

In the first hop, S transmits its packet to R over $N_{\rm sc}^{(1)}$ subcarriers with selected numerology $\mu^{(1)}$, mini-slot size $N_{\rm sym}^{(1)}$, and MCS $I_{\rm MCS}^{(1)}$. 
While the number of subcarriers $N_{\rm sc}^{(1)}$ varies depending on $\mu^{(1)}$, the same mini-slot size and MCS are applied across all subcarriers. 
Let $R_{C}^{(1)}$ and $M^{(1)}$ denote the coding rate and the modulation order corresponding to $I_{\rm MCS}^{(1)}$, respectively. 
The number of symbols required to send $H$ bits under the selected MCS $I_{\rm MCS}^{(1)}$ is given by $m_{1} =  \left \lceil H / \big( R_{C}^{(1)} \times M^{(1)} \big) \right \rceil$, where $ \lceil . \rceil$ is the ceiling function. 
The number of subframes required to carry $m_{1}$ symbols is given by \cite{URLLC_adpative_trans_2}
\begin{align} \label{eq:subframe}
N_{\rm sf}^{(1)} = \left \lceil \frac{m_{1}}{ N_{\rm sc}^{(1)} \times N_{\rm sym}^{(1)}} \right \rceil . 
\end{align}
The transmission time interval (TTI) for the first-hop transmission attempt is given by 
\begin{align} \label{eq:TTI}
t_{\rm TTI}(\mu^{(1)}, N_{\rm sym}^{(1)}, I_{\rm MCS}^{(1)}) = N_{\rm sf}^{(1)} \times T_{\rm sf}^{(1)}. 
\end{align}
For URLLC applications, the size of data packets is short and finite to comply with the stringent latency constraint \cite{FBL_relaying_1}.  
Thus, the assumption of infinite blocklength underlying Shannon's capacity theorem is no longer valid \cite{URLLC_relaying}. 
Under such a finite blocklength regime, the decoding error probability becomes non-negligible and must be considered in performance analysis and system design. 
Applying the finite blocklength regime result, the decoding error probability at R for the first hop can be approximated as \cite{FBL_error_rate}
\begin{align} \label{eq:decoding_error_pr}
\varepsilon_{1} (\gamma_{1}, m_{1}) = Q \left( \ln 2 \sqrt{\frac{m_{1}}{V_{1}}} \left( \log_{2} (1+ \gamma_{1} ) - \frac{H}{m_{1}} \right) \right), 
\end{align}
where $V_{1}=1-(1 + \gamma_{1})^{-2}$ is the channel dispersion for S--R link, and $Q\left( x \right) = (1/\sqrt{2 \pi}) \int_{x}^{\infty} \exp\left( -t^{2}/2 \right) dt $
is the Gaussian Q-function. 
If a decoding error occurs at R, an ARQ request is sent over one OFDM symbol to S \cite{URLLC_adpative_trans_1}, \cite{URLLC_adpative_trans_2}. 
Then, S selects the numerology, mini-slot size, and MCS to retransmit the packet. 
We assume that the ARQ request is always received successfully, and the overhead required to send an ARQ request is determined by the selected numerology \cite{URLLC_adpative_trans_1}.\footnote{Upon receiving an ARQ request, S reselects the numerology, mini-slot size, and MCS for retransmission. 
The number of OFDM symbols used in each (re)transmission attempt varies according to the newly selected MCS. }

In the second hop, upon successful packet reception over first hop, R selects its own numerology $\mu^{(2)}$, mini-slot size $N_{\rm sym}^{(2)}$, and MCS $I_{\rm MCS}^{(2)}$ to transmit the decoded packet to D over $N_{\rm sc}^{(2)}$ subcarriers. 
The number of symbols required to transmit $H$ bits over the second hop can be obtained by $m_{2} =   \left \lceil  H / \big( R_{C}^{(2)} \times M^{(2)}  \big) \right \rceil $, where $R_{C}^{(2)}$ and $M^{(2)}$ are the coding rate and the modulation order associated with $I_{\rm MCS}^{(2)}$. 
Similar to (\ref{eq:TTI}), the TTI for the second hop transmission can be calculated by 
\begin{align}
t_{\rm TTI}(\mu^{(2)}, N_{\rm sym}^{(2)}, I_{\rm MCS}^{(2)}) = N_{\rm sf}^{(2)} \times T_{\rm sf}^{(2)} ,
\end{align}
where $ N_{\rm sf}^{(2)}$ is the number of subframes required for second hop transmission and can be obtained by (\ref{eq:subframe}). 
The decoding error probability at D in the second hop is given by 
\begin{align} \label{eq:decoding_error_pr_2}
\varepsilon_{2} (\gamma_{2}, m_{2})= Q \left( \ln 2 \sqrt{\frac{m_{2}}{V_{2}}} \left( \log_{2} (1+ \gamma_{2} ) - \frac{H}{m_{2}} \right) \right) ,
\end{align}
where $V_{2}=1-(1 + \gamma_{2})^{-2}$ is the channel dispersion for R--D link. 
If a decoding error is detected at D, the ARQ request is immediately sent to R. 
The (re)transmission attempts continue until the packet is successfully delivered to D or the latency budget is exhausted, which declares \textit{packet loss}.

The total end-to-end transmission time consists of the TTIs for (re)transmissions and ARQ overheads. 
Let $i = 1$ and $i = 2$ denote agents S and R, respectively, corresponding to the hop in which they participate for packet transmission. 
The total transmission time $\mathcal{T}$ can be formulated by 
\begin{align} \label{eq:total_TTIs}
\mathcal{T} = \mathop{\sum_{i \in \{ 1, 2\}} \sum_{n}} \left( t_{{\rm TTI}, n}(\mu^{(i)}, N_{\rm sym}^{(i)}, I_{\rm MCS}^{(i)}) + t_{{\rm ARQ}, n}(\mu^{(i)}) \right), 
\end{align}
where $t_{{\rm TTI}, n}(\mu^{(i)}, N_{\rm sym}^{(i)}, I_{\rm MCS}^{(i)})$ and $t_{{\rm ARQ}, n}(\mu^{(i)})$ are the TTI and the ARQ response time in $n$-th (re)transmission attempt of agent $i \in \{ 1, 2\}$, respectively. 
Our goal is to maximize the probability of successful packet delivery over two hops by jointly optimizing numerology, mini-slot size, and MCS selection at S and R for each (re)transmission attempt under the latency constraint $\mathcal{T} \le T_{\rm th}$. 

\section{Dual-Agent Distributed MDP Formulation} 
The uncertainties of wireless channels introduce random decoding errors. 
As a result, the actual time spent in each hop transmission is stochastic, which makes the total transmission time $\mathcal{T}$ a random variable. 
As the distribution of $\mathcal{T}$ is challenging to formulate, the latency constraint $\mathcal{T} \le T_{\rm th}$ cannot be satisfied by traditional optimization techniques.\footnote{To apply supervised deep learning approaches, we must have pre-labeled data on what the best transmission parameters or actions should be for each system state in advance. However, such data is challenging to obtain for stochastic environments where decoding outcomes vary with random channel realizations. To overcome this limitation, we adopt reinforcement learning to enable S and R to learn optimal decisions on transmission parameters by directly interacting with the environment. }
To address this issue, we formulate the adaptive transmission design of the two-hop cooperative communication system as a Markov decision process (MDP) \cite{RL_book}. 
Acting as independent agents, S and R optimally select transmission parameters via a sequential decision-making process to maximize end-to-end reliability while satisfying the latency constraint $\mathcal{T} \le T_{\rm th}$. 
Based on the above discussion, the adaptive transmission design problem is formulated as the following MDP. 

\textbf{State space $\mathcal{S}^{(i)}$:  }
Each agent observes the instantaneous SNR of its own link $\gamma_{i}$ to adapt transmission parameters accordingly. 
Note that the decision at S affects not only its transmission in the first hop but also the remaining latency budget available for R to perform (re)transmissions in the next hop. 
Thus, additional information on the channel quality of the next hop is required to make latency-aware decisions. 
As the instantaneous channel state of the next-hop transmission cannot be obtained due to the limitations of the feedback channel, we assume that the average SNR of the next hop $\bar{\gamma}_{i+1}$ is available. 
Both agents must know the packet size $H$ and the remaining latency budget in the current transmission attempt, denoted by $\tau_{n}$. 
The state for agent $i$ in $n$-th (re)transmission attempt is given by $s_{n}^{(i)} = \left( \gamma_{i}, \bar{\gamma}_{i+1}, H, \tau_{n} \right) \in \mathcal{S}^{(i)}$ with a dimension of $4$. 
For the terminal hop ($i=2$), there is no next hop to deliver the packet; therefore, we set $\bar{\gamma}_{i+1} = \infty$. 
The decision process has two absorbing terminal states,  
\begin{itemize}
\item \textbf{Success}: If the packet is successfully transmitted to the intended receiver (R for the first-hop and D for the second-hop) within the latency budget, the process transitions to the \textbf{Success} state. 
\item \textbf{Failure}: The process terminates in the \textbf{Failure} state if the remaining latency budget is exhausted before successful transmission. 
\end{itemize}

\textbf{Action space $\mathcal{A}$: }
The action space $\mathcal{A}$ consists of the available resource configurations in 5G NR, including the numerology $\mu$, mini-slot size $N_{\rm sym}$, and MCS $I_{\rm MCS}$. 
The action space is defined by 
\begin{align} \label{eq:action_space}
\mathcal{A} =\big\{ ( \mu, & N_{\rm sym},  I_{\rm MCS} ) : \mu  \in \{0, 1, 2, 3, 4\}, N_{\rm sym} \in \{2, 4, 7, 14\}, I_{\rm MCS} \in \{1, \dots, 15 \} \big\}, 
\end{align}
where each action tuple $(\mu, N_{\rm sym}, I_{\rm MCS})$ has a dimension of $3$. 
In $n$-th (re)transmission attempt, the action selected by agent $i$ is given by $a_{n}^{(i)} = \left( \mu, N_{\rm sym}, I_{\rm MCS} \right) \in \mathcal{A}$, which directly affects the TTI and decoding error probability.

\textbf{Transition dynamics: }
The transition probability $\mathcal{P}(s_{n+1} | s_{n}, a_{n})$ is governed by the decoding error probability and the variation of latency budget, both of which determine whether a packet is successfully transmitted or further retransmissions are required. 
Given the selected action $a_{n}^{(i)}$, the remaining latency budget is updated by 
\begin{align}
\tau_{n+1} = \tau_{n} -  t_{{\rm TTI}, n}(\mu, N_{\rm sym}, I_{\rm MCS}) - t_{{\rm ARQ}, n}(\mu). 
\end{align}
For agent $i$, the probability of transitioning to the next state $s_{n+1}^{(i)}$ upon taking action $a_{n}^{(i)}$ in the current state $s_{n}^{(i)}$ depends on the occurrence of a decoding error and remaining latency budget $\tau_{n+1}$. 
Thus, the state transition is defined by
\begin{align} \label{eq:transition_two_hop}
\mathcal{P}(s_{n+1}^{(i)} |s_{n}^{(i)}, a_{n}^{(i)})  & =
\begin{cases}
\mathbf{1}\{\tau_{n+1} \ge 0 \} \times ( 1-\varepsilon_{i} ), & \text{if $s_{n+1}^{(i)}=$ \textbf{Success}}, 
\\
\mathbf{1}\{\tau_{n+1} < 0 \} \times \varepsilon_{i}, & \text{if $s_{n+1}^{(i)}=$ \textbf{Failure}} ,
\\
\mathbf{1}\{\tau_{n+1} \ge 0 \} \times \varepsilon_{i}, & \text{otherwise} ,
\end{cases}
\end{align}
where $ \mathbf{1}\{ \cdot \}$ is the indicator function, and $\varepsilon_{i}$ is the decoding error probability of agent $i$. 
If the transmission is successful within the latency budget, the process transits to the \textbf{Success} state with probability of $( 1-\varepsilon_{i} )$. 
If the remaining latency is exhausted, the process moves to the \textbf{Failure} state. 
Otherwise, (re)transmission continues as long as there is sufficient latency. 

\textbf{Reward $\mathcal{R}^{(i)}$: }
Since S is responsible for packet transmission over two hops, its action must be evaluated depending on whether there is sufficient latency budget left for the next hop transmission. 
However, S cannot directly observe the transmission outcome of the second hop because R makes independent transmission decisions. 
To address this issue, we use the delay outage rate (DOR) \cite{DoR} as a metric to estimate the likelihood of successful packet delivery in the next hop, given the remaining latency budget $\tau_{n+1}$. 
According to \cite{DoR}, the delivery time required for agent $i$ to send a packet of size $H$ bits can be defined by 
\begin{align}
T_{\rm delivery}^{(i)}  =  \frac{H}{W \log_{2} \left(1 + \gamma_{i} \right) }. 
\end{align}
As a delay outage occurs when the packet delivery time exceeds the latency budget $\tau_{n+1}$, the DOR over the remaining latency $\tau_{n+1}$ is expressed as
\begin{align} \label{eq:P_out}
\mathcal{P} \Big( T_{\rm delivery}^{(i)} > \tau_{n+1} \Big)  = 1 - \exp \left( \frac{-1}{\bar{\gamma}_{i}} \left( 2^{\frac{H}{ W  \tau_{n+1} }} -1\right) \right)
\triangleq \mathcal{P}_{\rm DOR} \left( \bar{\gamma}_{i}, \tau_{n+1} \right) . 
\end{align}
Based on the DOR, we define the reward for agent $i$ as
\begin{align} \label{eq:reward_two_hop}
\mathcal{R}_{n+1}^{(i)}   =
\begin{cases}
1-\mathcal{P}_{\rm DOR} \left( \bar{\gamma}_{i+1}, \tau_{n+1} \right) , &\text{if $s_{n+1}^{(i)} =$ \textbf{Success}}, 
 \\
-1 , &\text{if $s_{n+1}^{(i)}=$ \textbf{Failure}},
\\
-0.1 , &\text{otherwise} , 
\end{cases}
\end{align}
which gives less reward for actions that leave an insufficient latency budget $\tau_{n+1}$ for the next-hop transmission. 
For a successful terminal-hop transmission, the agent $i=2$ receives a reward of $1$, since the DOR for its next-hop transmission is zero with $\bar{\gamma}_{i+1} = \infty$. 
We impose a strong negative reward of $-1$ for entering the \textbf{Failure} state to discourage packet loss. 
A small negative reward of $-0.1$ is applied to suboptimal decisions that do not lead to the \textbf{Success} state but waste the remaining latency budget.\footnote{Both agents are penalized for actions that fail to reach the \textbf{Success} state yet consume the latency budget. Such a design encourages the agents to minimize unnecessary retransmissions. Thus, the number of retransmission attempts is implicitly optimized as the policy converges.  }
\section{Dual-Agent Reinforcement Learning Solution}
In two-hop cooperative transmission, two agents, S and R, make independent decisions. 
To facilitate the distributed learning, we propose a dual-agent reinforcement learning-based cooperative latency-aware transmission (DRL-CoLA) algorithm, where S and R learn the hop-specific transmission policies $\pi_{i}: s \mapsto a$, $i\in \{1, 2\}$, using only local observations and ARQ feedback. 
Since the formulated MDP involves a continuous state space and a discrete action space, we employ a deep Q-network (DQN) algorithm to approximate the Q-value function via deep neural networks (DNNs), i.e., $Q_{i}(s, a) \approx Q_{i}(s, a; {\boldsymbol \theta_{i}})$ \cite{Conventional_DQN}, where $\boldsymbol \theta_{i}$ is a parameter of the main Q-network for agent $i\in \{1, 2\}$. 
In each (re)transmission attempt, the agent observes its current state $s_{n}^{(i)}$ and selects the action $a_{n}^{(i)}$ following the $\epsilon$-greedy policy. 
In turn, the environment will respond by providing the reward $\mathcal{R}_{n+1}^{(i)}$ and the next state $s_{n+1}^{(i)}$. 
Such experience is stored as a transition tuple $e_{n} = \left(s_{n}^{(i)}, a_{n}^{(i)}, \mathcal{R}_{n+1}^{(i)}, s_{n+1}^{(i)} \right)$ in a replay buffer $\mathcal{B}_{i}$. 
When training, a mini-batch $\mathcal{M}_{i}$ of stored experiences is sampled from the replay buffer $\mathcal{B}_{i}$ and used to calculate the mean squared error (MSE) loss function, given by 
\begin{align} \label{eq:loss}
\mathcal{L}_{i}\left( \boldsymbol \theta_{i} \right) = \mathbb{E}_{e_{n} \sim \mathcal{M}_{i} } \Bigl[  \left( y_{n}^{(i)} -  Q_{i} (s_{n}^{(i)}, a_{n}^{(i)}; \boldsymbol \theta_{i} ) \right)^{2} \Bigr], 
\end{align} 
with the target value defined as $y_{n}^{(i)} = \mathcal{R}_{n+1}^{(i)} + \gamma  \max_{a^{\prime} \in \mathcal{A} } Q_{i}(s_{n+1}^{(i)}, a^{\prime}; \boldsymbol \theta^{-}_{i} )$, where $\boldsymbol \theta^{-}_{i}$ is a parameter of the target Q-network, and $\gamma$ is the discount factor with $0 \le \gamma \le 1$. 
The parameter of the main Q-network $\boldsymbol \theta_{i}$ is updated via the gradient descent method to minimize the loss function $\boldsymbol \theta_{i} \gets \boldsymbol \theta_{i} - \alpha \nabla_{\boldsymbol \theta_{i} } \mathcal{L}_{i} \left( \boldsymbol \theta_{i} \right)$, where $\alpha$ is the learning rate, and the gradient $\nabla_{\boldsymbol \theta_{i} } \mathcal{L}_{i}$ is calculated by
\begin{align}
\nabla_{\boldsymbol \theta_{i}} \mathcal{L}_{i}(\boldsymbol \theta_{i})
= \mathbb{E}\Bigl[\big(y_{n}^{(i)} - & Q_{i}(s_n^{(i)}, a_n^{(i)}; \boldsymbol \theta_{i})\big)  \times \nabla_{\boldsymbol \theta_{i}} Q_{i}(s_n^{(i)}, a_n^{(i)}; \boldsymbol \theta_{i}) \Bigr].
\end{align}
Every $E^{\prime}$ episode, the parameter of the target Q-network $\boldsymbol \theta^{-}_{i}$ is copied from that of the main Q-network \cite{Conventional_DQN}. 
After training is completed, each agent obtains the trained Q-network $\boldsymbol \theta_{i}^{*}$. 
The optimal policy is derived by selecting the action that maximizes the optimal Q-value function in each state \cite{RL_book}.  
The process of training the Q-network is presented in \textbf{Algorithm 1}. 
The proposed DRL-CoLa algorithm for two-hop cooperative relaying is presented in \textbf{Algorithm 2}. These proposed algorithms are presented in Appendix A.

\section{Simulation Results} \label{Simulation_results}
In this section, we provide the simulation results to evaluate the effectiveness of the proposed DRL-CoLa algorithm. 
The simulation setting for experiments is presented in Appendix B. 

\begin{figure}[ht] 
    \subfloat[Probability of packet loss versus $T_{\rm th}$. ]{\includegraphics[width=0.50\textwidth]
        {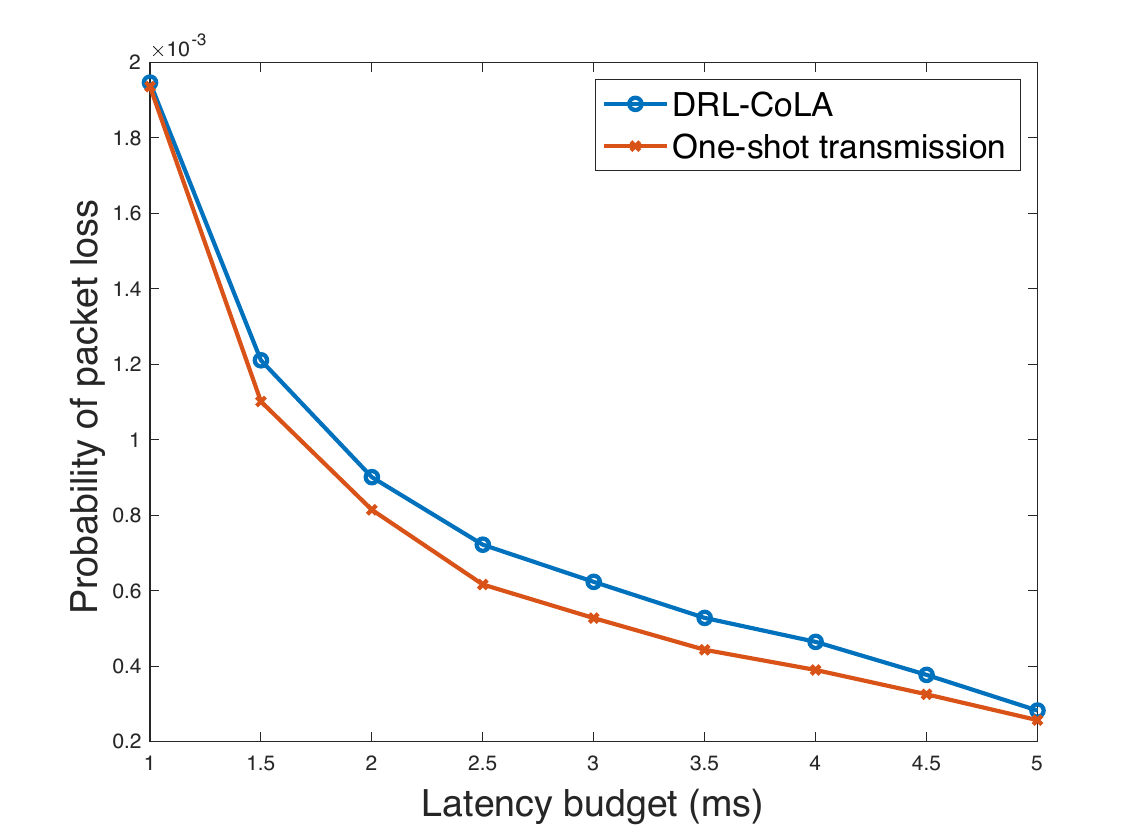}}
    \hfill
    \subfloat[Probability of packet loss versus $d_{1}$.]{\includegraphics[width=0.50\textwidth]
        {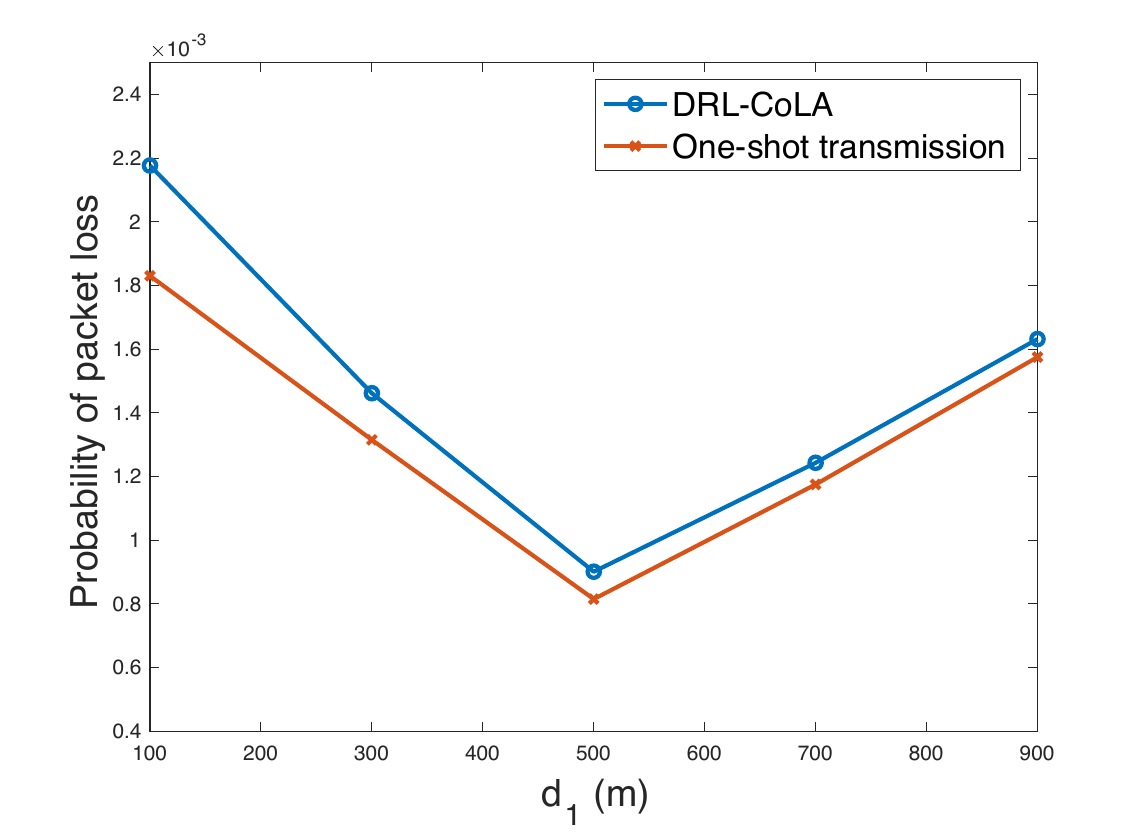}}
\caption{Comparison of end-to-end reliability between the proposed DRL-CoLa scheme and one-shot transmission. } 
\label{fig:figure_3}
\end{figure}

In Figure \ref{fig:figure_3} (a), we compare the proposed DRL-CoLA with the one-shot transmission scheme that optimally allocates symbols across both hops using global CSI. 
The one-shot transmission sets the lower bound on packet loss probability. 
Across the tested latency regimes, DRL-CoLA achieves near-optimal packet loss performance even without global CSI by learning hop-specific policies from local CSI and ARQ feedback. 
This result indicates that decentralized, per-hop decision making can satisfy stringent URLLC requirements while avoiding the overhead of global CSI acquisition.

In Figure \ref{fig:figure_3} (b), we plot the probability of packet loss versus the S--R distance $d_{1}$ while keeping $d_{1} + d_{2} = 1000$ m. 
The curve is V-shaped: probability of packet loss decreases as $d_{1}$ increases by bringing R closer to D, reaches a minimum at the symmetric placement $d_{1} = d_{2}$, and then increases once $d_{1}$ exceeds 500 m. 
This is because moving R toward D improves the link quality of R--D but simultaneously degrades that of S--R. 
Given the symmetric simulation parameters and identical fading statistics on both hops, balancing the large-scale losses $d_{1} = d_{2}$ leads to the lowest probability of packet loss. 
It is worth noting that packet loss is lower when $d_{1}>d_{2}$ than when  $d_{1}<d_{2}$. 
The reason is that the second-hop transmission is performed on a tighter latency budget than the first-hop transmission. 
Thus, improving the channel quality for the R--D link increases the likelihood of completing packet delivery within such a tight latency budget, thereby reducing the packet loss.

\begin{figure}[ht] 
    \subfloat[Rewards accumulated by S. ]{\includegraphics[width=0.50\textwidth]
        {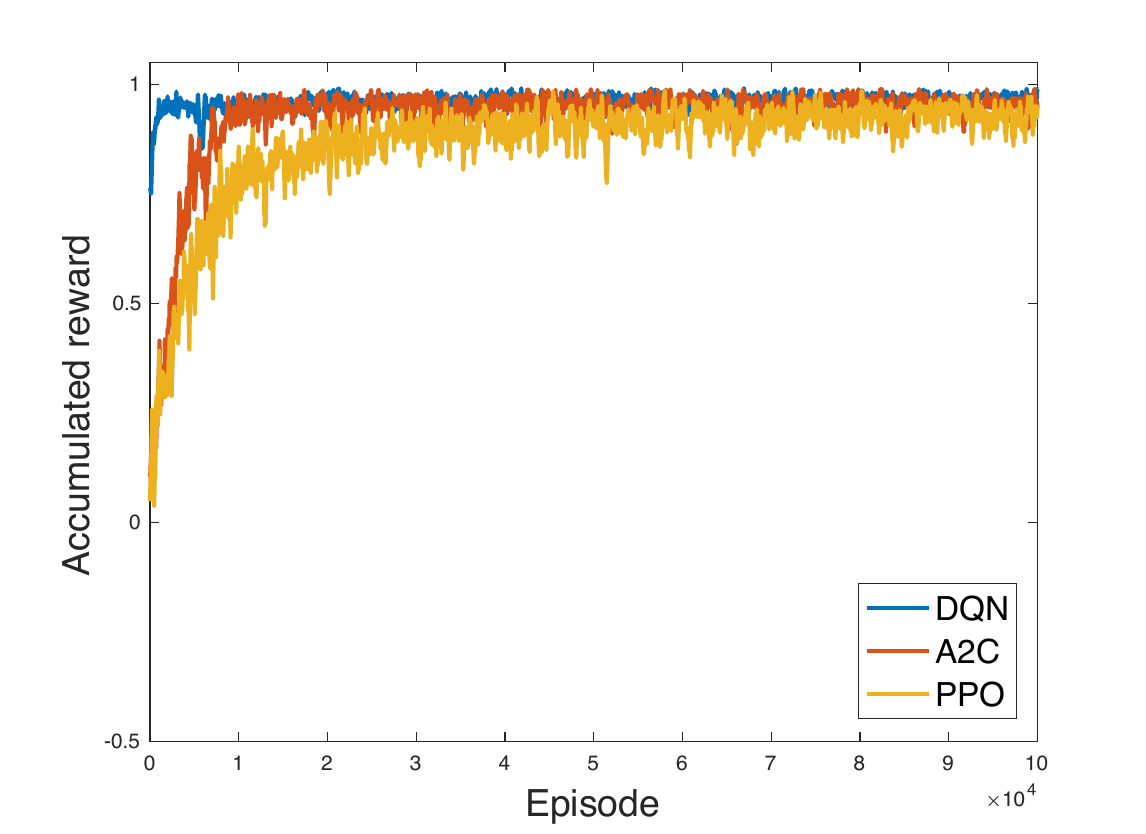}}
    \hfill
    \subfloat[Rewards accumulated by R. ]{\includegraphics[width=0.50\textwidth]
        {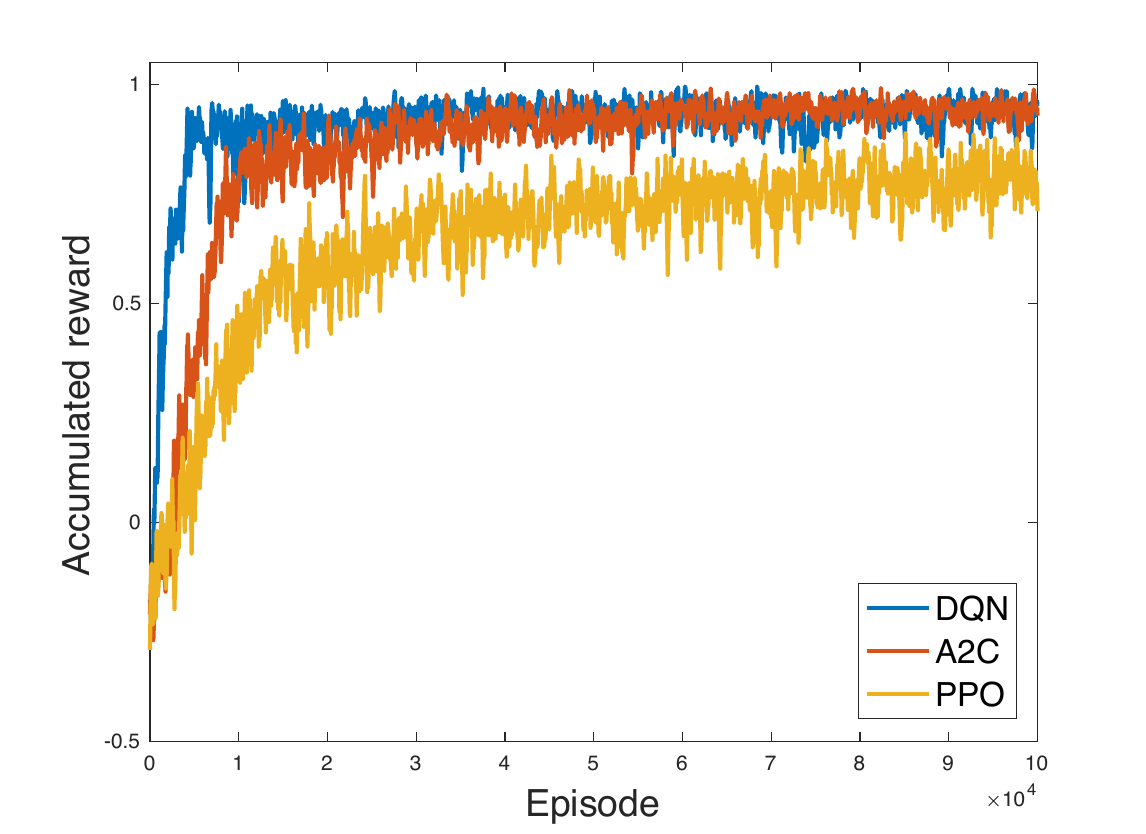}}
\caption{Comparison of accumulated rewards achieved by DRL-CoLA when implemented with different reinforcement learning algorithms.} 
\label{fig:figure_4}
\end{figure}

In Figure \ref{fig:figure_4}, we plot the accumulated rewards to demonstrate the convergence behavior of the proposed DRL-CoLA and to justify the selection of DQN for training both agents, S and R. 
We can see that the accumulated rewards of both agents increase steadily with training episodes and eventually stabilize. 
It confirms the convergence of the proposed DRL-CoLA algorithm. 
Moreover, DRL-CoLA trained with DQN shows faster convergence and achieves higher steady-state rewards compared to the A2C- and PPO-based implementation. 
This result indicates that the value-based DQN algorithm is more suitable for the considered cooperative transmission design, where a discrete action space allows DQN to learn near-optimal policies for both agents more efficiently than policy-gradient algorithms.

\section{Conclusion}
In this paper, we developed an adaptive transmission design for two-hop cooperative communication to meet the stringent URLLC requirements. 
We formulated the two-hop transmission process as an MDP to enable per-attempt radio resource configuration on the active hop for successful packet delivery across two hops within the latency budget. 
To solve the formulated MDP, we proposed the DRL-CoLA algorithm, where S and R learned decentralized, latency-aware policies from local observations and ARQ feedback.  
Simulation results showed that DRL-CoLA achieves near-optimal reliability comparable to the one-shot transmission scheme even without global CSI. 

\newpage
\bibliography{references}
\setcounter{section}{0} \setcounter{subsection}{0}
\setcounter{equation}{0}
\renewcommand{\theequation}{A.\arabic{equation}}
\section*{Appendix A\\Proposed Algorithms}

\begin{algorithm}[ht]
\caption{Training algorithm for Q-network}
\begin{algorithmic}[1]
\renewcommand{\algorithmicrequire}{\textbf{Input:}}
\renewcommand{\algorithmicensure}{\textbf{Output:}}
\REQUIRE Probability of exploration $\epsilon$; replay buffer $\mathcal{B}_{i}$; mini-batch size $|\mathcal{M}_{i}|$; initial state $s_{1}^{(i)}$;main Q-network parameter $\boldsymbol \theta_{i}$; target Q-network parameter $\boldsymbol \theta_{i}^{-}$
\ENSURE Remaining latency budget $T_{\rm rem}$; terminal flag $done^{(i)}$, updated replay buffer $\mathcal{B}_{i}$; updated main Q-network parameter $\boldsymbol \theta_{i}$
\STATE Set $done^{(i)} \gets$ False and $n \gets 1$. 
\WHILE{not $done^{(i)}$}
\STATE Observe the current state $s_{n}^{(i)}$ and select the action $a_{n}^{(i)}$ using the $\epsilon$-greedy policy. 
\STATE Receive the reward $\mathcal{R}_{n+1}^{(i)}$ and the next state $s_{n+1}^{(i)}$. 
\STATE Store transition $e_{n}^{(i)} = \left(s_{n}^{(i)}, a_{n}^{(i)}, \mathcal{R}_{n+1}^{(i)}, s_{n+1}^{(i)} \right)$ in the replay buffer $\mathcal{B}_{i}$.
\STATE Randomly sample a mini-batch $\mathcal{M}_{i}$ of transition tuples from the replay buffer $\mathcal{B}_{i}$. 
\STATE Update the main Q-network by performing gradient descent on loss $\mathcal{L}_{i} \left( \boldsymbol \theta_{i} \right)$ in (\ref{eq:loss}). 
\IF {$s_{n+1}^{(i)}$ = \textbf{Success} or $s_{n+1}^{(i)}$ = \textbf{Fail},}
\STATE $done^{(i)} \gets$ True. 
\ELSE
\STATE $n \gets n+1$. 
\ENDIF
\ENDWHILE
\STATE Set the remaining latency by $T_{\rm rem} \gets \tau_{n+1}$. 
\end{algorithmic}
\end{algorithm}

\begin{algorithm}[ht]
\caption{Proposed DRL-CoLa algorithm }
\begin{algorithmic}[1]
\renewcommand{\algorithmicrequire}{\textbf{Input:}}
\renewcommand{\algorithmicensure}{\textbf{Output:}}
\REQUIRE Number of episodes $E_{\rm max}$; probability of exploration $\epsilon$; epsilon decaying rate $\lambda$; frequency of updating target Q-network $E^{\prime}$; mini-batch size $|\mathcal{M}_{i}|$; packet size $H$; latency budget $T_{\rm th}$
\ENSURE Trained Q-network parameters $\boldsymbol \theta_{1}^{*}$ and $\boldsymbol \theta_{2}^{*}$
\STATE Initialize the main Q-network $\boldsymbol \theta_{i}$ and the target Q-network with $\boldsymbol \theta_{i}^{-} \gets \boldsymbol \theta_{i} $. 
\STATE Initialize the relay buffer with $\mathcal{B}_{i} \gets \emptyset$. 
\FOR{$e = 1, \cdots, E_{\rm max}$} 
\STATE // \textit{Perform first-hop transmission}
\STATE Initialize the state $s_{1}^{(1)}$ with packet size $H$ and latency budget $T_{\rm th}$. 
\STATE Invoke \textbf{Algorithm 1} to obtain the remaining latency $T_{\rm rem}$, first-hop termination status $done^{(1)}$, updated replay buffer $\mathcal{B}_{1}$, and parameter $\boldsymbol \theta_{1}$.  
\IF {$T_{\rm rem} \ge 0$ and $done^{(1)}=$ True,}
\STATE // \textit{Perform second-hop transmission}
\STATE Initialize the state $s_{1}^{(2)}$ with packet size $H$ and latency budget $T_{\rm rem}$. 
\STATE Invoke \textbf{Algorithm 1} to obtain the updated replay buffer $\mathcal{B}_{2}$ and parameter $\boldsymbol \theta_{2}$.  
\ENDIF
\STATE Decay exploration rate $\epsilon \gets \lambda \epsilon$. 
\STATE Every $E^{\prime}$ episode, update the parameters of the target Q-network by $\boldsymbol \theta_{1}^{-} \gets \boldsymbol \theta_{1} $ and $\boldsymbol \theta_{2}^{-} \gets \boldsymbol \theta_{2} $. 
\ENDFOR
\end{algorithmic}
\end{algorithm}
\setcounter{section}{0} \setcounter{subsection}{0}
\setcounter{equation}{0}
\section*{Appendix B\\Simulation Setting}
In this appendix, we detail the simulation settings used to generate the results in Section~\ref{Simulation_results}. 
We set $P_{1} = 30$ dBm, $P_{2} = 30$ dBm, and $\eta = 2$.  
The bandwidth is set to $W = 480$ kHz \cite{3GPP_numerology} with the noise power spectral density $N_{0} = 10^{-14}$ W/Hz (i.e., $-110$ dBm/Hz), which leads to the noise power as $\sigma^{2} = N_{0} \times W$.  
Unless otherwise specified, we set the latency budget as $T_{\rm th} = 2$ ms, the distance as $d_{1} = d_{2} = 500$ m, and $H = 256$ bits.
The main and target Q-networks are implemented as DNNs with three fully connected layers, consisting of $64$, $256$, and $128$ neurons, respectively, with \textit{ReLU} activations. 
Additionally, the Adam optimizer is employed to train both Q-networks. 
The hyperparameters for training \textbf{Algorithm 2} are summarized in Table \ref{tab:2}. 

\begin{table}[h]
\caption{Hyperparameters for \textbf{Algorithm 2}}
\label{tab:2}
\setlength\tabcolsep{0pt} 
\begin{tabular*}{\columnwidth}{@{\extracolsep{\fill}} ll cccc}
\toprule
     Parameter & Description  & Value \\ 
\midrule
     $E_{\rm max}$ & Number of episodes & $100000$  \\
     $E^{\prime}$ & Frequency of updating target & $2000$  \\
     $\gamma$ & Discount factor & $0.95$  \\
     $\alpha$ & Learning rate & $10^{-5}$  \\
     $\epsilon$  & Probability of exploration & $1$  \\
     $\lambda$ & Epsilon decaying rate & $0.999$  \\
     $\mathcal{B}_{i}$ &Size of replay buffer & $10000$ \\
     $\mathcal{M}_{i}$ &Size of mini-batch & $64$ \\
\bottomrule
\end{tabular*}
\end{table}

\end{document}